\documentclass[]{article}

\usepackage[version=3]{mhchem}
\usepackage{booktabs, multirow, dcolumn}
\usepackage{amssymb}
\usepackage{pdflscape}
\usepackage{caption,subcaption}
\usepackage{graphicx}
\usepackage{geometry}
\usepackage{epstopdf}
\usepackage{authblk}
\usepackage{rsc}
\usepackage{siunitx}

\usepackage{color}

\author[1]{Jeongjae Lee}
\author[2]{Bartomeu Monserrat}
\author[1]{Ieuan D. Seymour}
\author[1,3]{Zigeng Liu}
\author[2]{Si{\^a}n E. Dutton}
\author[1]{Clare P. Grey}

\affil[1]{Department of Chemistry, Lensfield Road, Cambridge CB2 1EW, United Kingdom}
\affil[2]{Cavendish Laboratory, J. J. Thomson Avenue, Cambridge CB3 0HE, United Kingdom}
\affil[3]{Current Address: Max Planck Institute for Chemical Energy Conversion, Stiftstr. 34–36, M{\"u}lheim an der Ruhr 45470, Germany}

\title{An \textit{Ab Initio} Investigation on the Electronic Structure, Defect Energetics, and Magnesium Kinetics in \ce{Mg3Bi2}}

\begin{document}
\maketitle

\begin{abstract}
We present a comprehensive \textit{ab initio} investigation on \ce{Mg3Bi2}, a promising Mg-ion battery anode material with high rate capacity. Through combined DFT (PBE, HSE06) and $G_0W_0$ electronic structure calculations, we find that \ce{Mg3Bi2} is likely to be a small band gap semiconductor. 

DFT-based defect formation energies indicate that Mg vacancies are likely to form in this material, with relativistic spin-orbit coupling significantly lowering the defect formation energies. We show that a transition state searching methodology based on the hybrid eigenvector-following approach can be used effectively to search for the transition states in cases where full spin-orbit coupling is included. Mg migration barriers found through this hybrid eigenvector-following approach indicate that spin-orbit coupling also lowers the migration barrier, decreasing it to a value of 0.34 eV with spin-orbit coupling. Finally, recent experimental results on Mg diffusion are compared to the DFT results and show good agreement. This work demonstrates that vacancy defects and the inclusion of relativistic spin-orbit coupling in the calculations have a profound effect in Mg diffusion in this material. It also sheds light on the importance of relativistic spin-orbit coupling in studying similar battery systems where heavy elements play a crucial role.
\end{abstract}

\section{Introduction}
Extensive research has been performed on lithium-ion battery (LIB) systems in the last few decades, resulting in their integration into electric vehicles and handheld devices.\cite{van_noorden_better_2014} However, supply restrictions, high cost, and safety limitations of LIBs have kindled research into battery systems using alternative metal ions. Recently, batteries based on multivalent ion chemistries such as \ce{Mg^{2+}}, \ce{Al^{3+}}, and \ce{Ca^{2+}} have gained attention due to their low cost and high volumetric densities, features important for large-scale applications such as electric vehicles and grid-based energy storage systems.\cite{canepa_odyssey_2017} Of these, Mg-ion batteries (MIBs) are an attractive system due to the high abundance of Mg in the Earth's crust and the large volumetric capacity of Mg metal. However, currently at least three factors hinder the application of working MIB systems: (1) absence of electrolytes compatible with high-voltage cathodes, (2) slow diffusion of divalent \ce{Mg^{2+}} ions in electrode materials, and (3) passivation of Mg metal surfaces due to electrolyte decomposition reactions and the concomitant difficulties in stripping and plating Mg.\cite{mohtadi_magnesium_2014}

While possible high-voltage MIB cathode candidates have been identified,\cite{canepa_odyssey_2017} it remains extremely challenging to identify an electrolyte that is suitable for use with both high voltage cathodes and Mg metal. This incompatibility of electrolytes that are suitable for high voltage cathodes with Mg metal and vice versa has been a problem in developing a working MIB system. Hence, a number of studies have focused on using an alternative anode material other than Mg to bypass this problem.\cite{mohtadi_magnesium_2014} In this respect, bismuth metal is a promising anode material with a low discharge voltage of 0.2 V \textit{versus} Mg metal.\cite{shao_highly_2014,arthur_electrodeposited_2012,murgia_insight_2015} Bismuth alloys with magnesium to form an intermetallic \ce{Mg3Bi2} phase, resulting in a theoretical capacity of 385 mAh/g. However, the most interesting feature of bismuth anode is that it displays fast Mg ion insertion and de-insertion, a feature not present in most other Mg-ion electrodes.\cite{ramanathan_porous_2016,shao_highly_2014} Recently, our group has used \ce{^{25}Mg} Nuclear Magnetic Resonance (NMR) spectroscopy to study \ce{Mg3Bi2} and reported fast exchange between the two crystallographically distinct Mg sites in this material; a dependence of the exchange rate and activation energy on the preparation condition (ball-milling and electrochemical synthesis) was also demonstrated.\cite{liu_insights_2017}

 As ion dynamics are properties intimately related to the defect chemistry (which in turn depends heavily on the preparation conditions) and ultimately the electronic structure, we present an in-depth \textit{ab initio} investigation on the \ce{Mg3Bi2} system, the end product of discharge process. The results are presented in four sections: in the first section, we discuss the structural characteristics of \ce{Mg3Bi2}, identifying the possible Mg diffusion pathways. Next, electronic structures and defect energetics are described. Due to the well known problem of band gap underestimation in semilocal DFT, we use three different methods (PBE, HSE06, and $G_0W_0$) to calculate the electronic structure. In particular, the Greens function based $G_0W_0$ method is used to give the most accurate value for the minimum band gap, overcoming the band gap problem in semilocal DFT. 
 
 Bismuth is a heavy element where relativistic spin-orbit coupling is expected to have a significant influence on the properties relevant to battery chemistry, as shown for lead in a recent \textit{ab initio} study of the lead-acid battery.\cite{ahuja_relativity_2011} Thus, calculations were performed with and without the explicit spin-orbit coupling Hamiltonian. Inclusion of spin-orbit coupling was found to reduce the band gap, defect formation energies, and also the Mg migration barrier which was calculated using the hybrid eigenvector-following approach. Using these results we are able to reconcile and rationalize previous reports on the Mg mobility in \ce{Mg3Bi2}, in which very different Mg mobilities were assumed to occur due to the different concentrations of Mg vacancies in the samples. We also demonstrate that the hybrid eigenvector-following method can be a very efficient approach for locating transition states in systems where spin-orbit coupling is likely to play an important role.

\section{Methodology}
\subsection{Computational Details}
All Density Functional Theory (DFT) calculations were performed with the VASP code\cite{kresse_ab_1993,kresse_ultrasoft_1999} employing the projector-augmented wave (PAW) method\cite{blochl_projector_1994}. Spin-polarized Perdew-Burke-Ernzerhof (PBE) and Heyd-Scuseria-Ernzerhof (HSE06) exchange-correlation functionals were adopted.\cite{perdew_generalized_1996,heyd_erratum:_2006} For the energy and force calculations, PAW pseudopotentials treating the Mg $3s$ and Bi $5d$ as valence states were used, with a plane-wave basis cutoff of 350 eV.  All lattice relaxations were performed with 1.3 times the ENMAX value as defined in the pseudopotential file. In addition, additional support grid was used for the evaluation of augmented charges (ADDGRID=.TRUE.). Self-consistent field (SCF) cycles were converged with an energy tolerance of $10^{-4}$ eV. Monkhorst-Pack $k$-point sampling of $<0.05$ \AA\textsuperscript{-1} was used in the Brillouin zone. For the density of states and band structure calculations, Mg $2p$ states were also treated as valence states in the PAW potential and an increased cutoff of 550 eV was used. Cellular relaxations  SCF cycles were converged to a $10^{-6}$ eV limit. $\Gamma$-centered $k$-point sampling of $<0.03$ \AA\textsuperscript{-1} was used.

Single-shot $G_0W_0$ calculations were performed with $250$ frequency grid points and a $360$ eV plane-wave cutoff for the response function calculations. HSE06 wavefunctions were used as starting wavefunctions for the $G_0W_0$ calculations. Band gap convergence with respect to the number of frequency grid points, number of empty bands, and plane-wave cutoffs were checked. Quasiparticle energy iterations were converged to a $10^{-8}$ eV limit. 


Relativistic corrections to the electronic structure were taken into account through two levels of theory: a `scalar relativistic' correction which only includes the mass-velocity and Darwin terms of the relativistic Hamiltonian, as implemented in VASP;\cite{hafner_ab-initio_2008} and a `full relativistic' correction that explicitly includes the spin-orbit coupling term alongside the scalar relativistic correction.

The initial structure of \ce{Mg3Bi2} was fully relaxed until the energy differences between the subsequent steps are converged to $10^{-5}$ eV per cell and the forces are $<0.05$ eV/\AA\textsuperscript{-1}. For the defect calculations, two supercells each containing 40 atoms (16 Mg, 24 Bi) and 120 atoms (48 Mg, 72 Bi) were created and again relaxed to the above criteria. Defect notations follow that of Kroger and Vink, with charge superscripts omitted for clarity (\textit{e.g.} Bi\textsubscript{Mg(oct)} denotes a neutral Bi sitting on an octahedral Mg site).\cite{kroger_relations_1956} All defect energies were referenced to the respective bulk metals (Mg, Bi) at the same level of theory as the defect calculations. Only the neutral defects are considered since the system has a vanishing band gap under the PBE level of theory, with the metallic behaviour enhanced under the inclusion of spin-orbit coupling. Whereas the formation energies under the HSE06 level of theory (where a finite band gap was observed) could give a chemical potential dependence, this was not possible due to the computational resources available.

\subsection{Transition State Searching}
Transition state and steepest-descent pathway were found using the hybrid eigenvector-following approach as implemented in the OPTIM code.\cite{seymour_mapping_2015} Details on the algorithm are available elsewhere.\cite{munro_defect_1999,kumeda_transition_2001} In brief, this method finds the smallest eigenvalue of the Hessian matrix without calculating the full matrix, by minimizing the Rayleigh-Ritz ratio. Here we use a low-memory Broyden-Fletcher-Goldfarb-Shannon (LBFGS) scheme to minimize this ratio on a given point to find the uphill path. The minimization was deemed to have occurred when the root-mean-square gradient is smaller than 0.025 eV \AA\textsuperscript{-1}. Then up to five LBFGS minimization steps are performed in the tangent space until the root-mean-square gradient is smaller than $10^{-3}$ eV \AA\textsuperscript{-1}. Initial guesses of the transition state structure were produced by removing a nearby Mg atom and putting the migrating Mg atom in the middle of the proposed diffusion path. The steepest descent pathway from the transition state is found by displacing the atoms by 0.1 \AA\ from the transition state in the parallel and antiparallel directions to the eigenvector. Local minima are then found by the LBFGS algorithm with energy convergence of $10^{-3}$ eV.

We note that the transition state geometry and eigenvectors obtained from the significantly cheaper scalar relativistic calculations could be re-used as an initial guess for the fully relativistic calculations. We have observed that only $<5$ force evaluations are typically needed for the relativistic calculations when this approach is taken, which significantly reduces the computational requirement. In addition, unlike the classical nudged elastic band (NEB) method where only a few points along the pathway are sampled, HEF method can sample the path at an arbitrary step size. This allows us to capture the finer details of the energy variance along the pathway.

\section{Results and Discussion}

\subsection{Crystal structure}
\begin{figure}[h]
 	\centering
 	\includegraphics[width=\textwidth]{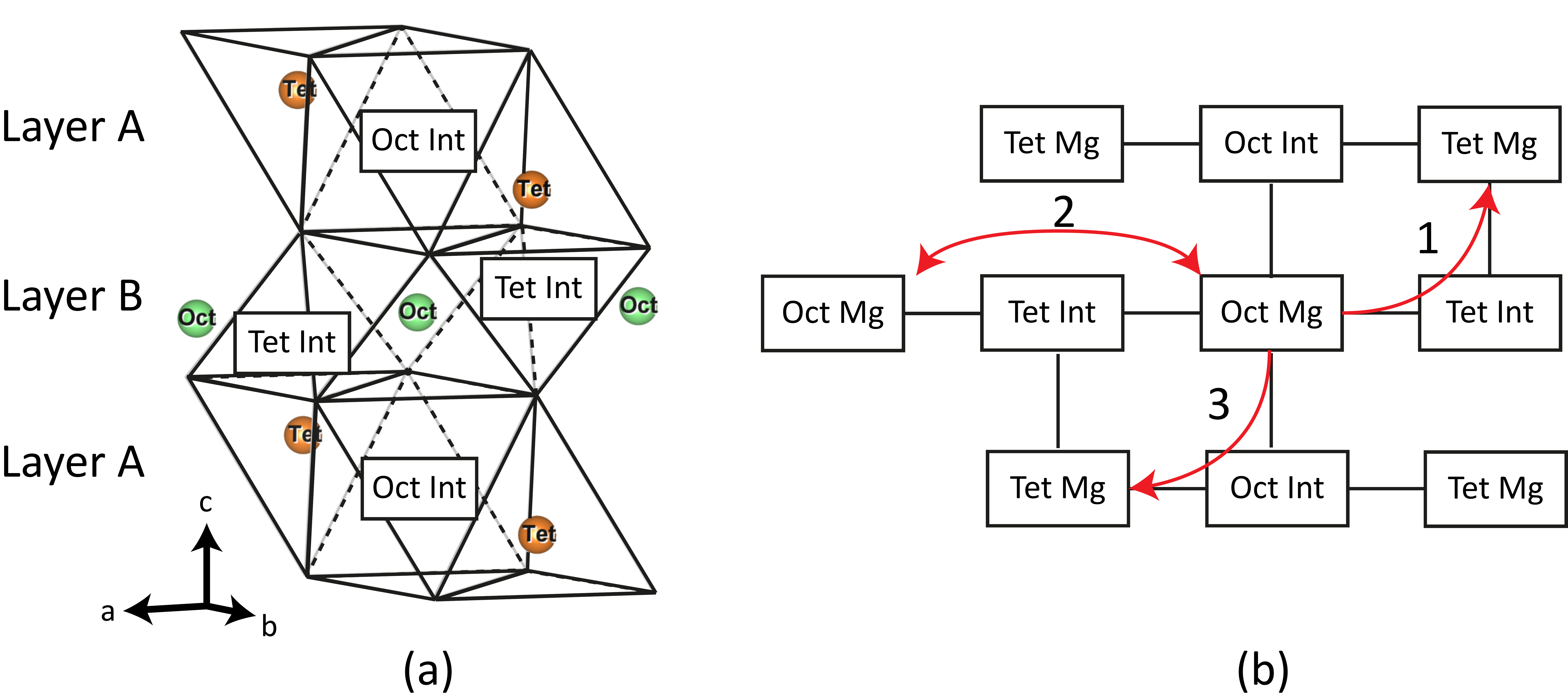}
 	\caption{(a) Crystal structure of \ce{Mg3Bi2}. Positions of tetrahedrally coordinated Mg (orange sphere, Tet) and octahedrally coordinated Mg (green sphere, Oct) are shown. Bismuth atoms sit on the line vertices (not shown for clarity). (b) Schematic illustration of the \ce{Mg3Bi2} structure (on the left) showing the possible diffusion pathways involving the tetrahedral and the octahedral interstitial sites (`Tet Int' and `Oct Int').}
 	\label{fig:structure}
 \end{figure}

\ce{Mg3Bi2} is the last in a series of magnesium pnictide compounds \ce{Mg3X2} (X=P, As, Sb, Bi) and is a Zintl-type compound with closed shell ions with formal charges of \ce{Mg^{2+}} and \ce{Bi^{3-}}. It adopts an anti-\ce{La2O3} type structure with hexagonal $P\bar{3}m1$ space-group symmetry.\cite{zintl_bindungsart_1933} The structure is schematically illustrated in Figure \ref{fig:structure}. In terms of ion arrangements, \ce{Mg3Bi2} incorporates two alternating layers: layer A consists of tetrahedrally coordinated  \ce{Mg^{2+}} cations and octahedral interstitial sites forming `covalent' \ce{Mg2Bi2^{2-}} sheets whereas layer B consists of octahedrally coordinated \ce{Mg^{2+}} cations and tetrahedral interstitial sites. These interstitial sites are expected to play an important role in ionic diffusion\cite{elliott_physics_1998} as discussed further below.
 
From DFT-based lattice relaxation, we can see that the PBE functional overestimates the experimentally determined $a$ cell parameter\cite{lazarev_preparation_1974} at both the scalar relativistic and full relativistic levels of theory, whereas it marginally underestimates the $c$ parameter at the scalar relativistic level, and overestimates it at the full relativistic level (Table \ref{tab:cell}). The HSE06 functional again overestimates the $a$ lattice parameter, although to a smaller extent than PBE, whereas it underestimates the $c$ lattice parameter. Finally, inclusion of SOC results in slight expansion of the cell which may arise from reduced electrostatic interactions (see below). 

\begin{table}[h]
	\centering
	\begin{tabular}{rrrrrr}
		& \multicolumn{2}{c}{PBE} & \multicolumn{2}{c}{HSE06}  & \\
		& SR & FR & SR & FR & Expt \\
		\cmidrule{1-6}
		$a$ / \AA & 4.71 & 4.72 & 4.65 & 4.66 & 4.62 \\
		$c$ / \AA & 7.40 & 7.44 & 7.35 & 7.36 & 7.41 \\
	\end{tabular}
	\caption{DFT-predicted cell parameters of \ce{Mg3Bi2} using the PBE and HSE06 exchange-correlation functionals. SR and FR refer to scalar relativistic and full relativistic (i.e. explicit spin-orbit coupling) calculations, respectively. Experimental lattice constants are from Lazarev \textit{et al}.\cite{lazarev_preparation_1974}}
	\label{tab:cell}
\end{table}

 Finally, we note that \ce{Mg3Bi2} undergoes a phase transition above 703 $^\circ$C to a defective body-centered cubic structure similar to \ce{AgI}, with excess Mg cations as the tetrahedral interstitials, resulting in \ce{Mg_{1.5}Bi}.\cite{barnes_fast-ion_1994} As expected from the AgI structure, it shows superionic conduction of \ce{Mg^{2+}} cations as determined through neutron diffraction.\cite{howells_quasielastic_1999} In this study, however, we restrict the investigation to the room-temperature hexagonal phase which is more relevant to Mg-ion batteries.

\subsection{Electronic structure}
Despite being first reported in 1933 by Zintl\cite{zintl_bindungsart_1933}, the precise electronic structure and the band gap ($E_g$) of \ce{Mg3Bi2} has not been determined with either experiment or theoretical calculations. In a series of work, Ferrier and co-workers have speculated that \ce{Mg3Bi2} is semimetallic based on a conductivity-composition plot;\cite{ferrier_conduction_1969,ferrier_conduction_1970} Lazarev \textit{et al}. have assumed \ce{Mg3Bi2} is semiconducting with $E_g=0.1$ eV;\cite{lazarev_preparation_1974} Watson \textit{et al}. obtained $E_g=0.5$ eV from Mg X-ray emission spectra.\cite{watson_electronic_1984} Work on the amorphous \ce{Mg3Bi2} alloy have shown from conductivity measurements that it is semiconducting with a band gap of 0.15 eV.\cite{sutton_filling-pseudogap_1975} These varying results may arise from the difficulty in preparing these materials stoichiometrically, due to the high vapor pressure of Mg. Previous literature describing electronic structure calculations is also sparse: Sedighi \textit{et al}. concluded \ce{Mg3Bi2} is a semiconductor with $E_g=0.25$ eV based on Engel-Vosko Generalized Gradient Approximation (EV-GGA),\cite{sedighi_density_2013} whereas Imai \textit{et al}., Xu \textit{et al}., and Zhang \textit{et al}. concluded it is a semimetal based on pure GGA calculations.\cite{imai_electronic_2006,xu_resistivities_1993,zhang_mg3bi2_nodal_line} However, all these works except the last one did not consider explicit spin-orbit coupling (SOC), which has been shown to play an important role in the electronic structure of compounds involving heavy atoms such as bismuth.\cite{shao_spin-orbit_2016,noh_spin-orbit_2008}; in addition, the well known problem of band gap underestimation in semilocal DFT necessitates the use of a higher level of theory to predict the accurate electronic structure of this material. Hence, we revisit the electronic structure of this compound using state-of-the-art electronic structure methods such as hybrid functionals and many-body perturbation theory in the $G_0W_0$ approximation, and also consider the effects of spin-orbit coupling.

First we look at the Bader charges of Mg and Bi in the structure, shown in Table \ref{tab:bader}. 
The HSE06 functional results in an increased ionicity in the system compared to the semilocal PBE values.
In all cases, inclusion of SOC resulted in decreased ionicity which is consistent with the density of states data shown below. 


\begin{table}[h]
	\centering
	\begin{tabular}{rrrr}
		& Bi    & Mg\textsubscript{oct} & Mg\textsubscript{tet}   \\
		\cmidrule{1-4}
		PBE SR   & -2.10 & +1.42        & +1.39          \\
		PBE FR      & -2.06 & +1.39        & +1.37          \\
		HSE06 SR & -2.21 & +1.51        & +1.45          \\
		HSE06 FR    & -2.18 & +1.48        & +1.44          \\
	\end{tabular}
	\caption{Scalar relativistic (SR) and full relativistic (FR) Bader charge analysis of \ce{Mg3Bi2} using the Perdew-Burke-Ernzerhof (PBE) and Heyd-Scuseria-Ernzerhof (HSE06) exchange-correlation functionals.
Only the valence charge is calculated.}
	\label{tab:bader}	
\end{table}

\begin{figure}
	\centering
	\includegraphics[width=\textwidth]{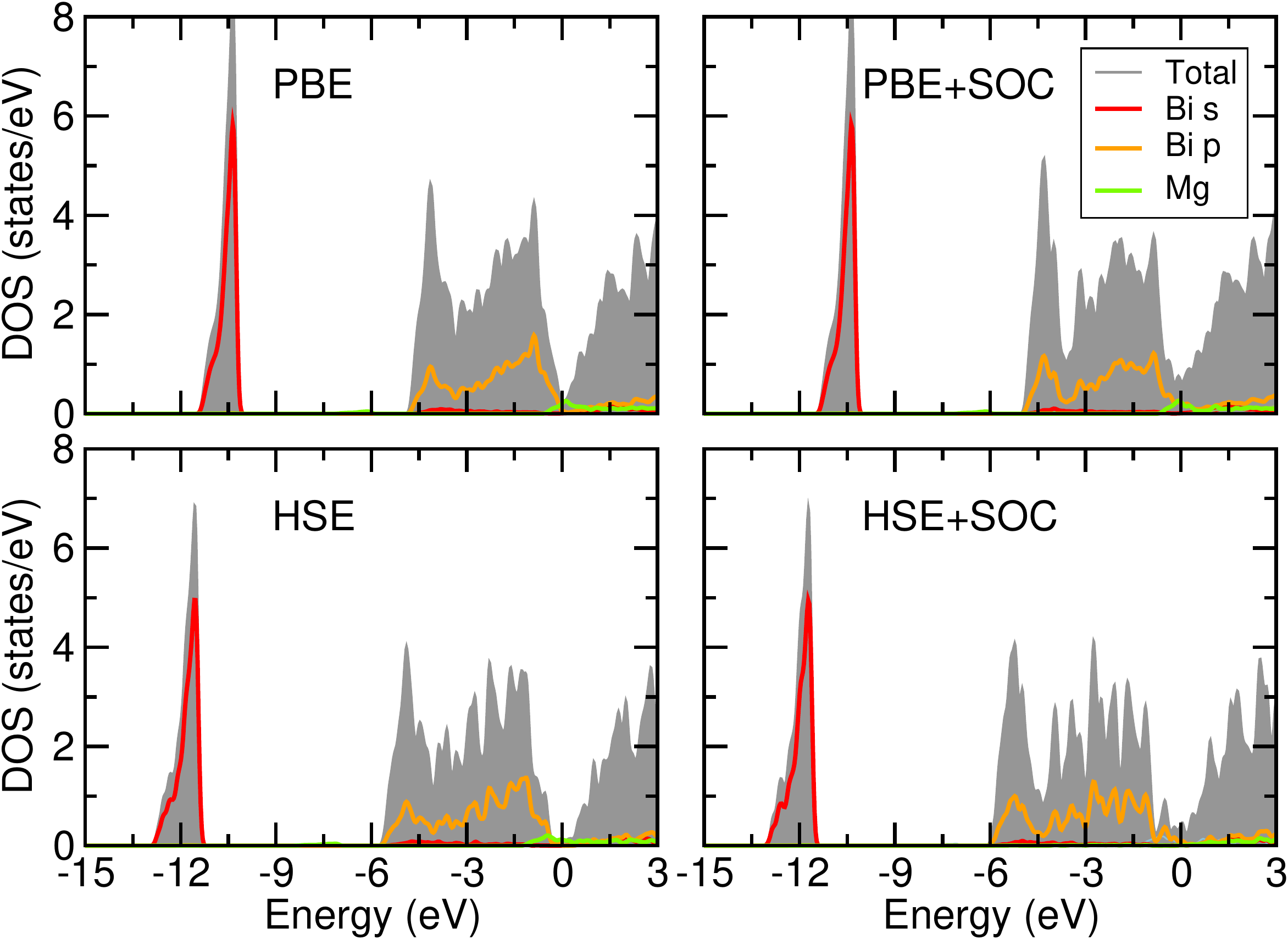}
	\caption{Density of states plot of \ce{Mg3Bi2} using the PBE and HSE06 exchange-correlation functionals. Local atomic DOS projections inside the sphere defined by the Wigner-Seitz radii (1.63 and 1.52 \AA\ for Bi and Mg, respectively) are also shown. All energies were referenced to the highest occupied state.}
	\label{fig:dos}
\end{figure}

Figure \ref{fig:dos} shows the density of states (DOS) plots obtained for \ce{Mg3Bi2} using DFT within the semilocal PBE and hybrid HSE06 approximations.  
Both methods show the same trend comparing the results with and without SOC. As expected from the charge distribution of \ce{Mg3Bi2}, the top of valence band is strongly dominated by the bismuth $6p$ contribution down to around $-5$ eV from the Fermi level and has a negligible Mg contribution. As expected from the literature, the relativistic contraction of the bismuth $6s$ states around $-12$ eV from the Fermi level results in a relatively large separation of around $5$ eV between the top of $6s$ and the bottom of $6p$ states, which makes it chemically inactive (the so-called inert-pair effect).\cite{yatsimirskii_relativistic_1995}. The downshift in $6s$ energy by adding the full relativistic effect is around -0.15 eV; effects of similar magnitudes were observed in \ce{PbO2}, an important active material in lead-acid batteries.\cite{ahuja_relativity_2011}

\begin{figure}
	\centering
	\includegraphics[width=0.45\textwidth]{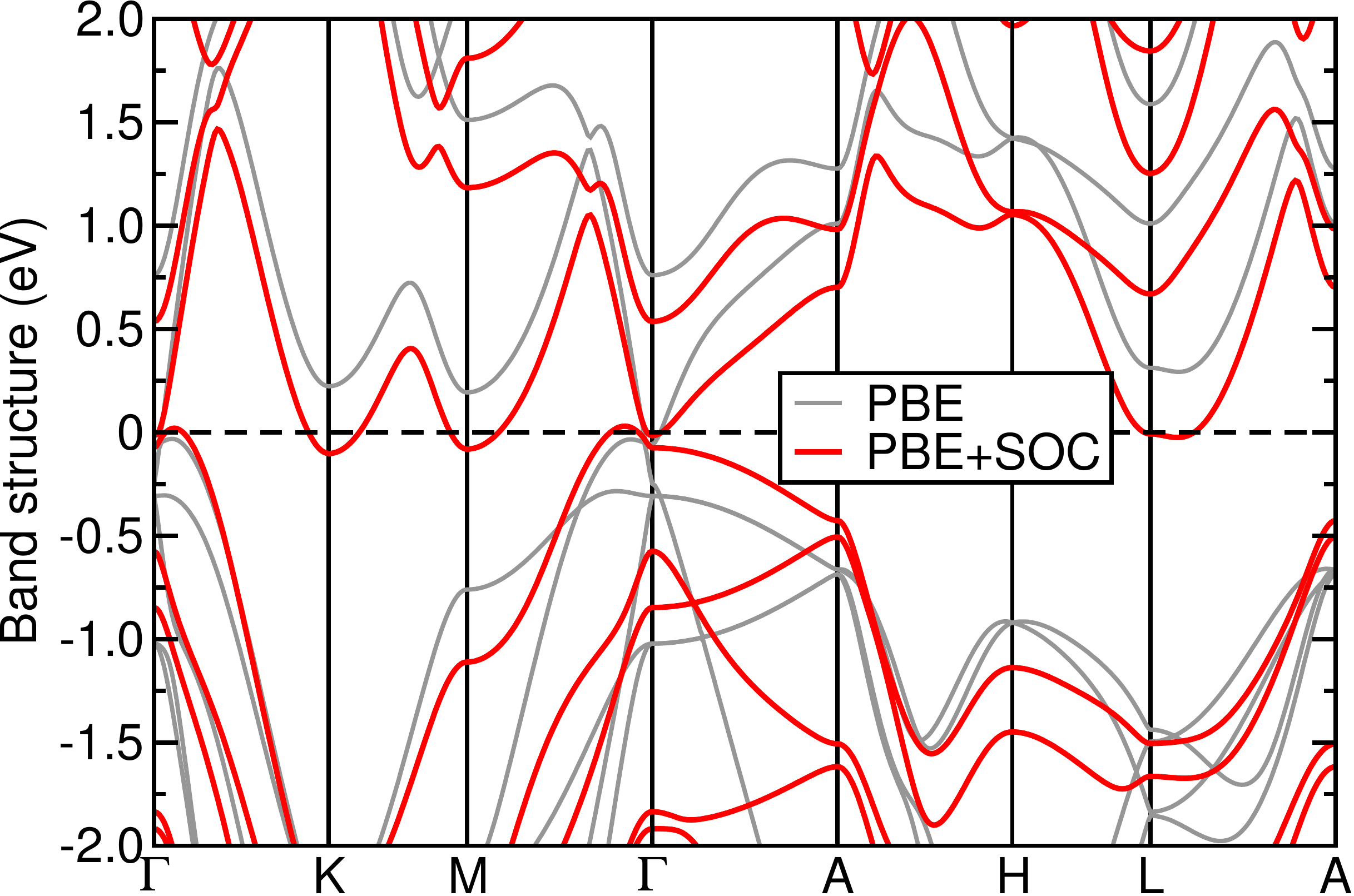}
	\includegraphics[width=0.45\textwidth]{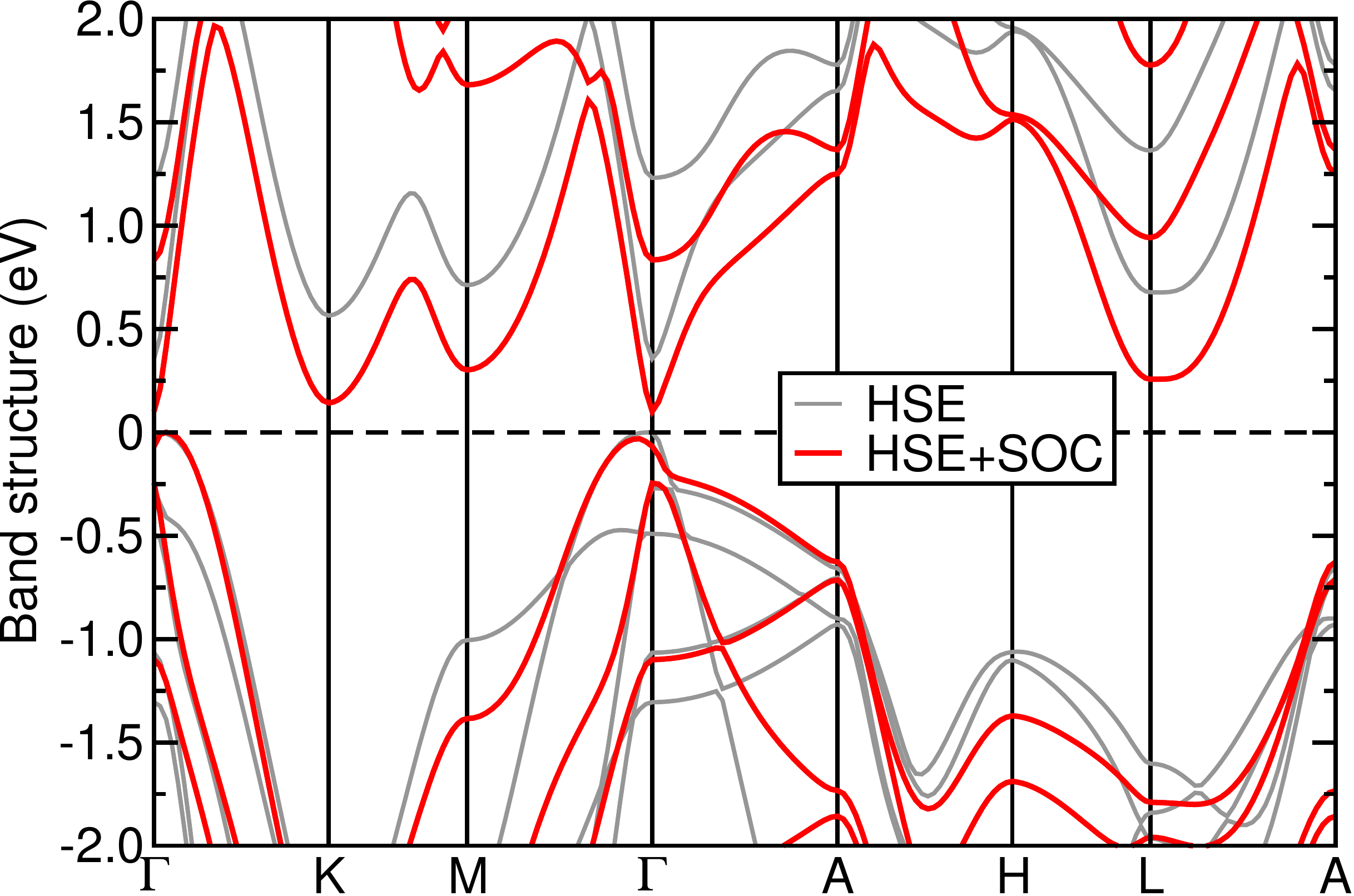}
	\caption{Band structure plots of \ce{Mg3Bi2} using the PBE (left) and HSE06 (right) exchange-correlation functionals without (grey) and with (red) spin-orbit coupling (SOC). The Fermi level is located at the zero of energy.}
	\label{fig:bs}
\end{figure}

Figure \ref{fig:bs} shows the band structure of \ce{Mg3Bi2} using DFT within the semilocal PBE and hybrid HSE06 approximations, with and without the spin-orbit interaction. At the PBE level, \ce{Mg3Bi2} exhibits semimetallic behavior, consistent with earlier reports of a type-II nodal line semimetal at this level of theory.\cite{zhang_mg3bi2_nodal_line} The inclusion of spin-orbit coupling opens a small gap in the nodal line, but the system remains semimetallic. At the HSE06 level, \ce{Mg3Bi2} is a semiconductor with the minimum band gap of $0.36$ eV located at the $\Gamma$ point. The inclusion of spin-orbit coupling reduces the band gap to $0.17$ eV, but the system remains semiconducting. The $G_0W_0$ calculations show that the $\Gamma$-point band gap increases to $0.58$ eV without spin-orbit coupling. We expect that spin-orbit coupling would reduce the band gap by an amount similar to that observed in HSE06, and thus \ce{Mg3Bi2} is also semiconducting at the $G_0W_0$ level of theory.


We note that inclusion of SOC results in a reduced band gap and increased dispersion of the Bi $6p$ states. Previous reports on binary Sb and Te compounds have attributed this phenomenon to spin-orbit coupling in the anion $p$-orbitals: energy splitting of the degenerate $p$-states result in the $j=3/2$ state being split upward in energy, and the $j=1/2$ state being split downward in energy.\cite{crowley_resolution_2016} The net effect of this splitting is a reduction in the band gap. 

\subsection{Defect energetics}
To investigate the possibility of different types of defects present in the \ce{Mg3Bi2} sample, formation energies $\Delta E_f$ of various stoichiometric and non-stoichiometric point defects were calculated \textit{ab initio}. The results outlined in Table \ref{tab:defect} are separated into three categories: antisite defects, Frenkel defects, and vacancy defects, which we discuss in sequence. Nonstoichiometric antisite and vacancy defects were created by removing or inserting relevant atoms from the stoichiometric cell while retaining the charge neutrality of the cell due to \ce{Mg3Bi2} being a semimetal at the PBE level of theory (\textit{i.e.} charged defects are not possible under such a situation). For instance, a Mg vacancy defect would involve removal of a neutral Mg atom resulting in a composition of \ce{Mg23Bi16} (for a 40-atom supercell) and two electron holes on the Fermi level; similarly, a Mg substitutional defect would involve a Bi atom replaced for a Mg atom, resulting in \ce{Mg25Bi15} (again for a 40-atom supercell) and five excess electrons on the Fermi level. Charge neutrality is automatically maintained in the antisite and Frenkel defects since they are stoichiometric.

First, the formation of antisite defects are energetically unfavorable, with the Mg on a Bi site having the highest formation energy of 2.92 eV. This is most likely due to the large difference between their ionic radii. Despite the fact that no ionic radius for the \ce{Bi^{3-}} ion is reported, the trend can still be explained in terms of the corresponding atomic radii of Mg and Bi (Mg 1.50 \AA; Bi 1.60 \AA), the difference being expected to get even larger as Mg is oxidized and Bi is reduced. This is also supported from the Mg--Bi bond lengths in \ce{Mg3Bi2} crystal: Mg\textsubscript{oct}--Bi 3.21 \AA \ \textit{versus} Mg\textsubscript{tet}--Bi 2.92 \AA. Using the Shannon radii of 0.72 and 0.57 \AA \ for Mg\textsubscript{oct} and Mg\textsubscript{tet}, respectively,\cite{shannon_revised_1976} \ce{Bi^{3-}} radii of 2.49 and 2.35 \AA \ are obtained in each case. In addition, the unfavourable electrostatic interaction between the ions of same charge (Mg occupying the Bi lattice site is coordinated by Mg ions, and vice versa) reinforces the high energy cost in forming these types of defects.

The $\Delta E_f$ of Frenkel-type defects (creation of a vacancy plus an interstitial) are also shown to be relatively high for both nearby and separated vacancy--interstitial pairs; around 0.8-1.2 eV is required for their formation. However, performing structural relaxation on some of the starting guesses with the defect Mg ion sitting on an interstitial site that is adjacent (in the first coordination shell) to a Mg vacancy (specifically, the $\mathrm{V_{Mg(oct)}+Mg_{i(tet)}}$ (nn), $\mathrm{V_{Mg(tet)}+Mg_{i(tet)}}$ (nn), and $\mathrm{V_{Mg(tet)}+Mg_{i(tet)}}$ (nn) cases) resulted in the structure reverting back to that of the pristine cell, with no defects. This indicates that these defects are energetically unstable and they would revert back to the original structure, if formed at all. These sites, as we will see later, may play an important role as energy maximum transition state sites in Mg-ion diffusion.

Perhaps the most surprising results are the energies of the vacancy defects: while Mg vacancy defects have around the same $\Delta E_f$ as Frenkel defects without SOC, their magnitudes are reduced significantly when SOC is included in the Hamiltonian. From the fully relativistic calculation on the 120-atom supercell with SOC included, $\Delta E_f$ of octahedral vacancies $\mathrm{V_{Mg(oct)}}$ is found to be as low as 0.33 eV, with a slightly higher value of 0.42 eV for tetrahedral vacancies $\mathrm{V_{Mg(tet)}}$. This effect is likely to be connected to the enhanced shielding of Bi $6p$-levels due to the SOC effect as explained above: as formation of a neutral Mg vacancy should involve loss of electrons from the Bi, this deshielding should result in a significant reduction of the formation energy. Hence, considering their low formation energies, we conclude that vacancy defects are the dominant type of defects present in \ce{Mg3Bi2}.

\begin{table}[]
	\centering
	\begin{tabular}{lrrrr}
		& \multicolumn{4}{c}{$\Delta E_f$} \\
		& \multicolumn{2}{c}{40-atom Supercell} & \multicolumn{2}{c}{120-atom Supercell} \\
		& SR & FR                & SR & FR  \\
		\cmidrule{1-5}
		\multicolumn{5}{c}{Antisite defects}      \\
		$\mathrm{Bi_{Mg(oct)}}$                   & 1.84   &                    &        &      \\
		$\mathrm{Bi_{Mg(tet)}}$                   & 2.01   &                    &        &      \\
		$\mathrm{Mg_{Bi}}$                       & 2.92   &                    &        &      \\
		$\mathrm{Bi_{Mg(oct)}+Mg_{Bi}}$          & 2.51   & 2.27               &        &      \\
		$\mathrm{Bi_{Mg(tet)}+Mg_{Bi}}$          & 1.17   & 1.15               &        &      \\
		\multicolumn{5}{c}{Frenkel defects}      \\
		$\mathrm{V_{Mg(oct)}+Mg_{i(oct)}}$ (nn)  & 0.87   & 0.79               & 0.89       &      \\
		$\mathrm{V_{Mg(oct)}+Mg_{i(oct)}}$ (far) & 0.89   & 0.81               & 1.04       &      \\
		$\mathrm{V_{Mg(oct)}+Mg_{i(tet)}}$ (nn)  & *      & *                  & *       & *     \\
		$\mathrm{V_{Mg(oct)}+Mg_{i(tet)}}$ (far) & 1.49   & 1.41               & 1.55       &      \\
		$\mathrm{V_{Mg(tet)}+Mg_{i(oct)}}$ (nn)  & *      & *                  & *       & *    \\
		$\mathrm{V_{Mg(tet)}+Mg_{i(oct)}}$ (far) & 1.07   & 0.98               & 1.10       &      \\
		$\mathrm{V_{Mg(tet)}+Mg_{i(tet)}}$ (nn)  & *     & *                  & *       & *     \\
		$\mathrm{V_{Mg(tet)}+Mg_{i(tet)}}$ (far) & 1.17   & 1.15               & 1.61       &      \\
		\multicolumn{5}{c}{Vacancy defects}      \\
		$\mathrm{V_{Mg(oct)}}$                   & 1.09   & 0.40                  & 1.03   & 0.33 \\
		$\mathrm{V_{Mg(tet)}}$                   & 1.17   & 0.50                & 1.12   & 0.42 \\
		$\mathrm{V_{Bi}}$                        & 1.88   & 2.02                &        &     
				\end{tabular}
		\caption{Scalar relativistic (SR) and full relativistic (FR) \textit{ab initio} formation energies $\Delta E_f$ of various stoichiometric and non-stoichiometric defects in \ce{Mg3Bi2} using the PBE functional. Defect notations follow the convention of Kroger and Vink with neutral sign omitted for clarity.\cite{kroger_relations_1956} All calculations assumed non-charged defects (see text) with cell dimensions fixed to simulate a dilute limit. Vacancy defect energies are referenced to the respective metals. For Frenkel defects, two scenarios where the Mg sits on a nearby (nn) or far interstitial sites were considered. Asterisks(*) indicate that the resulting structure was unstable and reverted back to the starting structure. All values are quoted in electron-volts. Only some of the calculations were performed under the 120-atom supercell condition after an initial screening with the 40-atom supercell.}
		\label{tab:defect}
			\end{table}
		
\subsection{Mg migration kinetics}
Having established the defect chemistry in \ce{Mg3Bi2}, we now turn our attention to the Mg diffusion in this structure and study the migration barriers.

As Mg vacancies were shown to have low formation energy, especially with SOC included, we attempt to simulate the effect of vacancy diffusion following the creation of one octahedral Mg vacancy $\mathrm{V_{Mg(oct)}}$. This has lower $\Delta E_f$ than the tetrahedral vacancy $\mathrm{V_{Mg(tet)}}$ as illustrated in the previous section. Then, a nearby tetrahedral Mg atom is removed from its original position and placed on the guessed `transition state'. Finally, a hybrid eigenvector-following approach is used to find the transition state , followed by the search for a steepest descent path connecting the two corresponding minima.

\begin{figure}
	\centering
	\includegraphics[width=\textwidth]{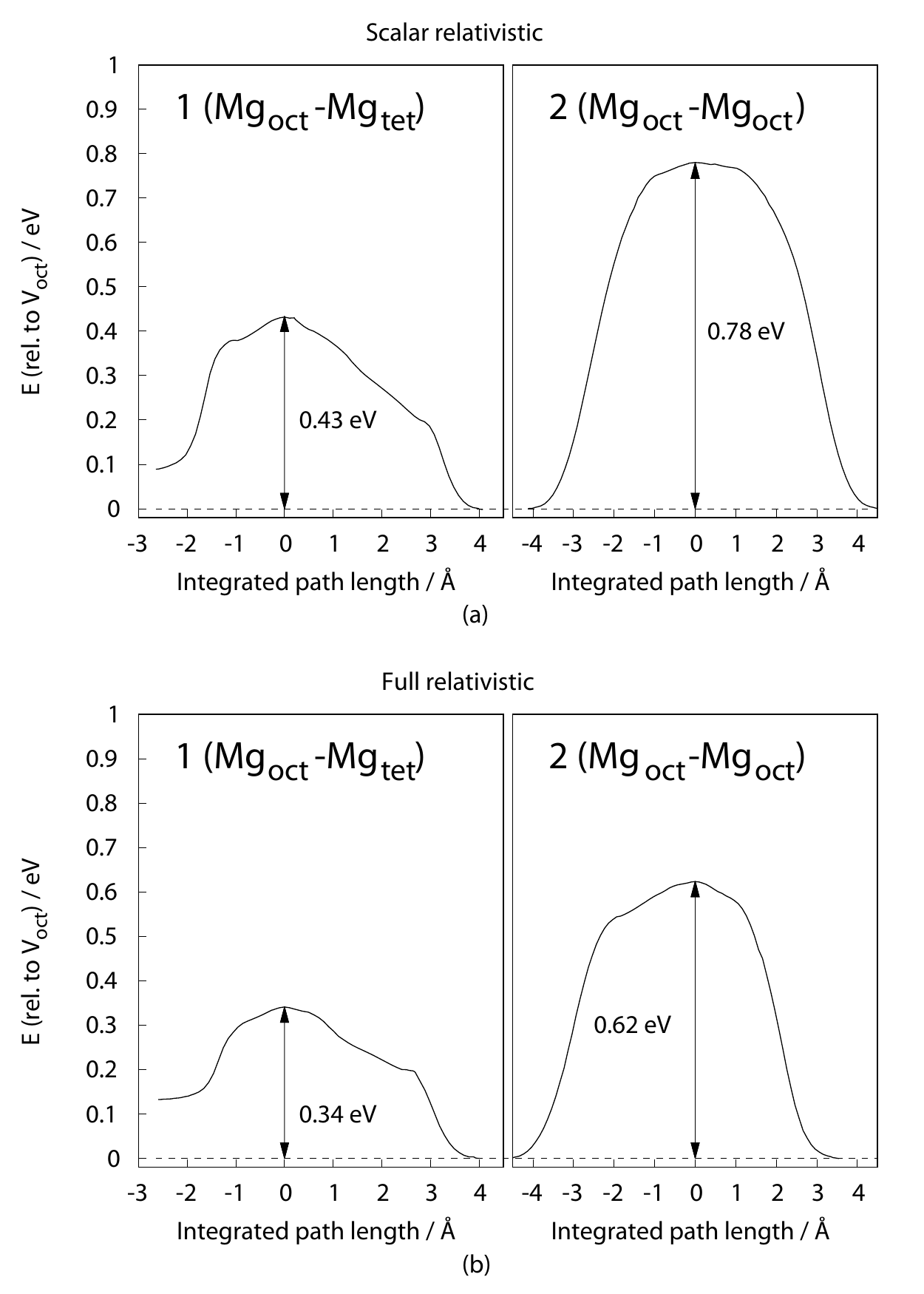}
	\caption{Diffusion profile of Mg ion for selected pathways as illustrated in Figure \ref{fig:structure}b. (a) Scalar relativistic, (b) full relativistic, with SOC included. Energies are referenced to the bulk energy of each 120-atom supercell with one $\mathrm{V_{oct}}$ defect.}
	\label{fig:optim}
\end{figure}

The result presented in Figure \ref{fig:optim} clearly shows a small diffusion barrier of hopping between the $\mathrm{Mg_{oct}}$ and $\mathrm{Mg_{tet}}$ (Path 1 on Figure \ref{fig:structure}b), with 0.34 eV for the fully relativistic calculation and 0.43 eV for the scalar relativistic calculation. In contrast, the $\mathrm{Mg_{oct}}-\mathrm{Mg_{oct}}$ diffusion barrier (Path 2 on Figure \ref{fig:structure}b) is around twice that of the $\mathrm{Mg_{oct}}-\mathrm{Mg_{tet}}$ diffusion barrier, indicating that the Mg diffusion must occur via octahedral-tetrahedral exchange. This is in line with the conclusion from \ce{^{25}Mg} NMR studies of Liu \textit{et al}.\cite{liu_insights_2017} An alternative exchange mechanism involving Path 3 on Figure \ref{fig:structure}b was also investigated, but the transition structure searching resulted in the same transition state as for the Path 1, clearly indicating the absence of a diffusion pathway along this line.

 With the \textit{ab initio} calculations of the defect creation and activation energies, we now compare our results with available experimental data. As reported in the literature, bismuth can be cycled reversibly to form \ce{Mg3Bi2} in a Mg-ion battery.\cite{shao_highly_2014} Previous work on magnesium ion conduction in bismuth anodes either used the Galvanostatic Intermittent Titration Technique (GITT) on an electrochemical cell, or \textit{ex-situ} variable-temperature (VT) \ce{^{25}Mg} NMR spectroscopy to probe Mg transport.\cite{ramanathan_porous_2016,liu_insights_2017} In this work, however, we restrict the discussion to the latter NMR result, due to the following limitation of GITT experiments: in the GITT measurement, the voltage response (resulting from the relaxation) of the cell after the application of a short current pulse is modelled with the diffusion equation to extract the diffusion coefficient $D$ under the assumption that the (de)insertion reaction occurs via solid solution (i.e. the relaxation of the potential after the current pulse is a measure of ionic transport through a \textit{single} phase). In-situ X-ray diffraction results, however, have clearly shown that magnesiation occurs via a two-phase reaction between Bi and \ce{Mg3Bi2}.\cite{liu_insights_2017} Thus the diffusion coefficients extracted from GITT measurements must be treated with caution, since the relaxation phenomena under these conditions may include multiple contributions such as (i) redistribution of Mg ions due to the formation of metastable (possibly non-stoichiometric solid solution) phases formed under operating conditions (and on application of an overpotential), and (ii) redistribution of phase boundaries to minimise the interfacial energies, etc.

We now consider transport in two cases, the first in a completely stoichiometric material where we need to consider the energy associated with defect formation. In this case, substitutional and vacancy defects are not relevant because they result in stoichiometry changes. The second case explores diffusion in off-stoichiometry materials \ce{Mg_{3-x}Bi2} and \ce{Mg3Bi_{2+y}}, representing Mg vacancy and (excess) Bi substitutional defects, respectively. As no extrinsic Mg vacancies are expected in the stoichiometric \ce{Mg3Bi2} compound, the diffusion in this case must occur via a vacancy diffusion mechanism involving thermally generated Mg vacancies; hence the observed $D$ would be the diffusion coefficient of Mg vacancies. In the latter case, nonstoichiometry dictates that extrinsic Mg vacancies must exist in the compound.

Table \ref{tab:diffusion} compares the effective activation barriers $E_a^\mathrm{eff}$ obtained experimentally \textit{versus} the $E_a^\mathrm{eff}$ estimated from the \textit{ab initio} calculations using the scalar relativistic (SR) and full relativistic (FR) treatments. Looking at the SR case first, a large variation in $E_a^\mathrm{eff}$ is observed where a significantly lower barrier is predicted for the Mg jumps with an existing nearby extrinsic vacancy defect (i.e. Case 2, \ce{Mg_{3-x}Bi2} SR): 0.43 eV. In Case 1 (stoichiometric \ce{Mg3Bi2}), where generation of a thermal Mg vacancy is required for diffusion, the effective diffusion barrier $E_a^{\mathrm{eff}}$ should now include this formation of vacancy. Naturally, we would expect an increase in $E_a^\mathrm{eff}$ due to this inclusion of vacancy formation energy: depending on the vacant site (tetrahedral or octahedral Mg vacancy), the $E_a^\mathrm{eff}$ is estimated to be 1.46 eV (octahedral) or 1.55 eV (tetrahedral).

We also observe that inclusion of spin-orbit coupling enhances the diffusion noticeably in all three cases, where a reduction of effective activation barrier is seen (e.g. for Case 1, SR with $\mathrm{V_{oct}}$ 1.46 eV \textit{versus} FR $\mathrm{V_{oct}}$ 0.67 eV. This is mainly due to the vacancy formation energy, where the FR case results in a dramatically lowered formation energy.

Comparing these results to the experiment, the effective activation barrier $E_a^\mathrm{eff}$ in the fully relativistic case ignoring the vacancy creation (Case 2) agrees with the $E_a^\mathrm{eff}$ of the ball-milled sample, determined through VT NMR. On the other hand, $E_a^\mathrm{eff}$ obtained for the electrochemically prepared samples (VT NMR Echem) agree well with the $E_a^\mathrm{eff}$ assuming a Mg vacancy creation plus Mg diffusion (Case 1). Hence, we conclude that the primary diffusion mechanism in electrochemically prepared samples (measured \textit{ex-situ}, i.e. not during battery operating conditions) should involve vacancy creation, whereas the mechanism in mechanically prepared samples only involves the vacancy diffusion. This could be explained by method of sample preparation: the electrochemical Mg insertion process in this case produced samples closer to the thermodynamic equilibrium creating fewer vacancies, whereas mechanical milling is largely a high energy process resulting in more vacancies (for instance, vacancy formation in ZnO through milling was previously observed with HRTEM studies\cite{chen_influence_2014}). This is also a known phenomenon in the synthesis of intermetallic phases: mechanical milling can provide excess energy to the material, which can be stored in the sample as atomistic disorders of which vacancies are one example.\cite{suryanarayana_chapter_1999} Furthermore, as discussed above, Mg deficiency is likely due to preferential Mg sticking to the ball-mill components (jar, balls). The already present vacancies in mechanically prepared samples act as potential diffusion sites for adjacent Mg ions, enhancing their diffusion. 

Extending this result to a battery under operating conditions, it is important to stress that the \ce{Mg_{3-x}Bi2} phases formed \textit{in-situ} may not be stoichiometric. The kinetics of Bi (de)magnesiation will depend on a number of factors which include the interfacial energies between the Bi and \ce{Mg3Bi2} phases, and transport of Mg in Bi, \ce{Mg3Bi2}, and at the various interfaces. In particular, the mechanism of demagnesiation will depend strongly on the ease of vacancy formation in \ce{Mg3Bi2}. By analogy with our work on lithium silicides\cite{ogata_revealing_2014} and of others (e.g. \ce{CuTi2S4}\cite{yu_designing_2014}), this energy of vacancy formation may even be responsible for setting (or strongly influencing) the overpotential observed on charge.

Finally, we note that Zintl-type \ce{A_3B_2} materials to which \ce{Mg3Bi2} belongs have been identified as potentially promising thermoelectric materials.\cite{ponnambalam_thermoelectric_2013,sun_thermoelectric_2017} Mg mobility in these materials could have important implications on the thermoelectric power generation, and work is in progress on doping other atoms into this structure to enhance, or suppress, this mobility.

\begin{table}
	\centering
	\begin{tabular}{cccc}
		\multirow{2}{*}{Case} & \multirow{2}{*}{Diffusion process} & \multicolumn{2}{c}{$E_a^\mathrm{eff}$ / eV} \\
		& & SR & FR \\
		\cmidrule{1-4}
		\multicolumn{4}{c}{DFT} \\
		Case 1 & formation and diffusion of $\mathrm{V_{oct}}$ & 1.46 & 0.67 \\
		Case 1 & formation and diffusion of $\mathrm{V_{tet}}$ & 1.55 & 0.76 \\
		Case 2 & pre-formed vacancy diffusion& 0.43 & 0.34 \\
		\multicolumn{4}{c}{Experiment} \\
		& VT NMR Echem & \multicolumn{2}{c}{0.71}  \\
		& VT NMR Ball-mill & \multicolumn{2}{c}{0.19} \\
	\end{tabular}
	\caption{Effective migration barriers $E_a^\mathrm{eff}$ estimated through DFT and VT NMR techniques. VT NMR data are taken from Liu \textit{et al}.\cite{liu_insights_2017} For the NMR measurements, \ce{Mg3Bi2} prepared through electrochemical insertion and mechanical milling were considered. SR and FR refer to scalar relativistic and full relativistic calculations, as described in the text.}
	\label{tab:diffusion}
\end{table}

\section{Conclusion}
In conclusion, advanced electronic structure calculations show that spin-orbit coupling plays an important role in structure and dynamics of \ce{Mg3Bi2}, a promising Mg-ion battery anode material. Inclusion of relativistic spin-orbit coupling also significantly lowers the formation energies of Mg vacancy defects, which is crucial to the apparent low Mg-ion diffusion barrier. Using an efficient single-ended hybrid eigenvector-following approach, we have calculated the Mg migration barriers involving relativistic spin-orbit coupling which are as low as 0.34 eV for the octahedral to tetrahedral diffusion. The calculated activation barriers are in good agreement with the previous experimental report using variable temperature \ce{^{25}Mg} NMR experiments for materials prepared electrochemically and via ball-milling. Stoichiometric materials show higher activation energies, since the activation energy involves both the cost of vacancy generation and transport. An understanding of Mg transport and the energetics of vacancy formation are important in understanding the mechanisms for demagnesiation of \ce{Mg3Bi2}. This work sheds light on ways of improving the Mg diffusion in similar materials such as Sn, which have been shown to have good capacity but poor rate performance in Mg-ion batteries.\cite{singh_high_2013}

\section{Acknowledgement}
We thank Professor David Wales for helpful discussions. Via our membership of the UK's HEC Materials Chemistry Consortium, which is funded by EPSRC (EP/L000202), this work used the ARCHER UK National Supercomputing Service (http://www.archer.ac.uk). Research was also carried out at the Center for Functional Nanomaterials, Brookhaven National Laboratory, which is supported by the U.S. Department of Energy, Office of Basic Energy Sciences, under Contract No. DE-AC02-98CH10886. J.L. acknowledges Trinity College Cambridge for the graduate studentship. B.M. acknowledges support from the Winton Programme for the Physics of Sustainability, and from Robinson College, Cambridge, and the Cambridge Philosophical Society for a Henslow Research Fellowship. 

\section{Supporting Information}
Research data in support of this manuscript will be made available on Apollo, the University of Cambridge repository.

\bibliographystyle{rsc}
\bibliography{refs}

\end{document}